\documentclass[letterpaper,11pt]{article}

\usepackage{amsmath, amsthm, amssymb}

\usepackage{amsfonts}
\usepackage{graphicx}
\usepackage{epsfig}
\usepackage{color}
\usepackage{algorithm}
\usepackage[noend]{algorithmic}
\usepackage{fullpage}
\usepackage{graphics}
\usepackage{psfig}
\usepackage{subfigure}

\newtheorem{theorem}{Theorem}

\newtheorem{lemma}[theorem]{Lemma}

\newcommand{\Z}{{\cal Z}}
\newcommand{\eps}{{\epsilon}}
\def\myangle#1{\mbox{angle}(#1)}

\title{Quasi-Polynomial Time Approximation Schemes for Target Tracking}
\author{Matt Gibson \and Gaurav Kanade \and Erik Krohn \and Kasturi Varadarajan}

\begin{document}
\bibliographystyle{plain}
\maketitle

\section{Introduction}
Target Tracking is the problem of keeping track of a set of specified targets by means of a given set of sensors. We study the target tracking problem in which the targets lie in the plane and the sensors are cameras also positioned in the plane. It requires two distinct cameras to estimate the position of a target. The quality of this estimation depends mainly on the relative position of the target with respect to that of the two cameras assigned to it \cite{hz00,gmsvw08}. The field of view of a camera is a cone. A target can be tracked by a camera if it lies in this cone, and therefore a target tracked by a pair of cameras should lie in the intersection of their respective cones.

Tracking a target in this manner cannot in general provide accurate estimates of position. Hence it is important to carefully pick and assign pairs of cameras to different targets so as to minimize the error in estimation. With this setting our problem can be viewed as a resource allocation problem. For arbitrary error functions, this problem is NP-hard and hard to approximate - belonging to the class of \textit{Multi-Index Assignment Problems} \cite{s00}, but usually the error is some function of the geometry of the camera and target positions \cite{ikst05}. Other NP-hard versions of multi-index assignment problems also focus on geometry, such as those aiming to minimize the circumference or the area of a triangle formed by three assigned points in the plane \cite{sw96}. 

The problem we thus consider is the \textit{Focus of Attention} problem (FoA) which requires us to find a pairing of cameras and an assignment of camera pairs to a target in a manner that is optimum for some measure of tracking quality \cite{ikst05,gmsvw08}. In our work we shall consider the constrained geometric setting in which the cameras are stationed on a line in the plane. Past work on this problem has also focused on this constrained setting \cite{ikst05,gmsvw08} and it is likely to model well the scenario in which targets are at relatively large distances from viable camera positions. Although the cameras cannot move they can rotate and freely choose their viewing direction. Both the cameras and the targets are represented by points.

\begin{figure}
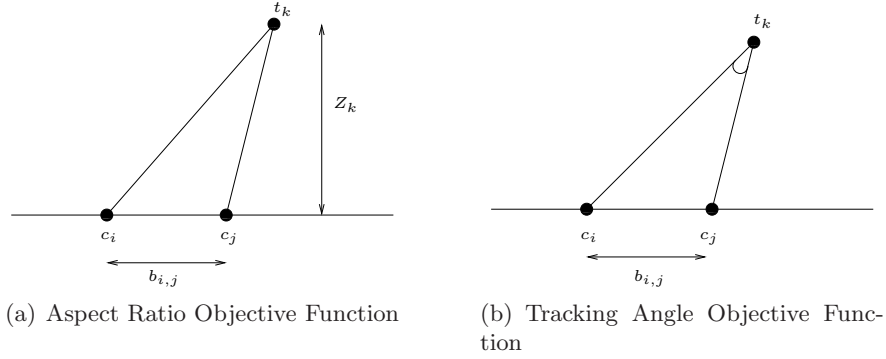

\centering
\subfigure[Aspect Ratio Objective Function]{
 \input{obj1.pstex_t}

\label{obj1}
}
\hspace{0.3in}
\subfigure[Tracking Angle Objective Function]{
 \input{obj2.pstex_t}
\label{obj2}
}

\caption{Here cameras $c_i$ and $c_j$ that are $b_{i,j}$ apart are assigned to target $t_k$ which is at a distance of $Z_k$ from the camera line}
\label{obj}
\end{figure}

We consider two geometric error metrics. The first is the ``Aspect Ratio'' which is the ratio of the vertical distance of the target from the camera line to the distance between the two cameras dedicated to it. (Refer Figure \ref{obj1}.) This metric can be used to gauge the error in stereo reconstruction and gives a good approximation if the cameras are not too close to the target. It was first considered by \cite{ikst05}. 

The second metric is the ``Tracking Angle Deviation From Right Angles'', studied by Gfeller et al. \cite{gmsvw08}, who state that for a pair of cameras tracking
a target, the tracking accuracy is best if the angle at the target 
- the \textit{tracking angle} - is closest to $90^\circ$. Thus the considered 
metric is the deviation of the angle from this desired value. (Refer Figure \ref{obj2}.) If the targets are not too close to the cameras, we can assume a scenario in which the tracking 
angles are all small i.e. less than $90^\circ$. In such a scenario minimizing 
the deviation from the optimum tracking angle is equivalent to maximizing the 
tracking angle. In this paper, we assume that all tracking angles are less 
than $90^\circ$ and hence our objective is to maximize the tracking angles.

We can assume without loss of generality that all the targets lie on one side 
of the line on which the cameras are placed. If they are not, we can consider 
their projections without affecting either of our optimization metrics.

Formally, we define the two versions of the FoA problem we consider as 
follows:\\
INPUT: A set $T$ of $n$ targets given as points in the plane and a set $C$ of $2n$ cameras, given as collinear points on line $l$ in the same plane.\\
FEASIBLE SOLUTION: A camera assignment where each target is assigned to two cameras and each camera is assigned to exactly one target.\\
MEASURE: (1) An Aspect Ratio and (2) A Tracking Angle for every triple consisting of a target and two cameras.\\
GOAL: Find a feasible solution which is optimal for the sum of (1) aspect ratios and (2) tracking angles.

\begin{figure}
\begin{center}
\input{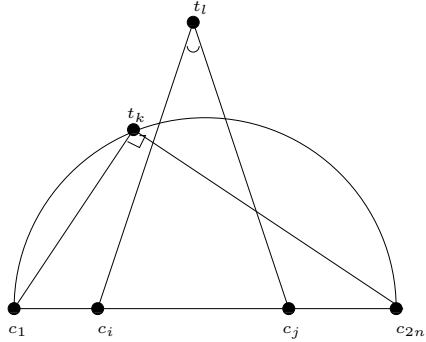}
\caption{Tracking angles are all less than $90^\circ$ if targets are not too close}
\label{thales}
\end{center}

\end{figure}

In this paper we consider the problem of (1) minimizing the sum of aspect ratios - the  MINSUMOFRATIOS problem and (2) maximizing the sum of tracking angles - the MAXSUMOFANGLES problem. For this second problem,
we assume that the input has the property that the tracking angle of
{\em every} triple is at most $90^\circ$, that is, every target lies outside 
the Thales' circle of any possible camera pair (Figure \ref{thales}.)

\paragraph{Related Work.}
Target Tracking is an important research topic in the field of computer vision 
and image processing and has applications in environment surveillance and 
monitoring applications \cite{yjs06,gmsvw08}.

Isler et al. \cite{ikst05} first studied target tracking by formulating the \textit{Focus of Attention} (FoA) problem as a combinatorial optimization problem. The motivation behind their work was lowering the costs of optimum depth estimation. They showed that in a general setting (not in the plane) this comprises the classical NP-Hard \textit{$3$-Dimensional Matching} (3DM) problem as a special case. Therefore the focus in \cite{ikst05} is on the constrained geometric setting in which all cameras are restricted to lie on a single line $l$. 
The objective is the ``aspect ratio'' $\frac{Z_k}{b_{i,j}}$ where $Z_k$ is the distance of target $t_k$ from $l$ and $b_{i,j}$ is the distance between cameras $c_i$ and $c_j$ assigned to target $t_k$ (also called the baseline). They give a $2$-approximation for the problem of minimizing the sum of aspect ratios 
(that is, the MINSUMOFRATIOS problem) and for the problem of minimizing the 
maximum aspect ratio. Also, if the cameras are placed equidistantly on the 
line, they present a PTAS for the MINSUMOFRATIOS problem. They also consider 
cameras on a circle and targets inside the circle with tracking cost being $\frac{1}{\sin{\theta}}$, where $\theta$ is the tracking angle, and deliver a $1.42$-approximation for the problem of minimizing the the sum of tracking costs, and the maximum tracking cost.

 Gfeller et al. \cite{gmsvw08} show that the problem of minimizing the sum of the deviations of tracking angles from $90^\circ$ (best tracking angle for accuracy) is NP-Hard, and that it admits no (multiplicative) approximation. For cameras on a line, they present a $2$-approximation algorithm for the problem of maximizing the sum of tracking angles (that is, MAXSUMOFANGLES) and maximizing the 
minimum tracking angle (or the bottleneck angle) under the assumption that all tracking angles are less than $90^\circ$. Also, if the cameras are placed equidistantly on the line, they present a PTAS for the MAXSUMOFANGLES problem. Arkin and Hassin \cite{ah97} give a $2+\frac{1}{t}$ approximation for the problem of maximizing the sum of tracking angles with cameras lying on a line (Here, $t$ is the size of the local neighborhood in the local-search algorithm).

\paragraph{Our Contribution and Techniques.} 
We consider the FoA problem that asks for camera assignment with minimum sum of Aspect Ratios (MINSUMOFRATIOS) and FoA that asks for camera assignment with maximum sum of tracking angles (MAXSUMOFANGLES). For cameras on a line, (and targets not lying in the Thales' circle of any camera pair) we present a Quasi-PTAS for 
MAXSUMOFANGLES. For cameras on a line, we present a Quasi-PTAS also for
the MINSUMOFRATIOS problem. A Quasi-PTAS is an algorithm that, for any constant
$0 < \epsilon < 1$, returns a solution whose cost is within an additive $\epsilon$ factor of the optimal and has time complexity $n^{\textrm{polylog}(n)}$. Thus we improve
upon the constant factor approximations known for these two problems,
using quasi-polynomial rather than polynomial time.
 
It is evident that the powerful geometric structure underlying the Focus of Attention Problem sets it apart from the more general assignment problems and makes it interesting. This geometry has been exploited in various ways by \cite{ikst05,gmsvw08} to obtain efficient approximation algorithms. In particular in both these efforts the cameras are divided into two sets of equal size and classified as ``left'' cameras and ``right'' cameras. 
Also in the special case of cameras placed equidistantly on the line both \cite{ikst05,gmsvw08} use the technique of further partitioning both the left and right cameras and then guessing the number of camera pairs in the optimal solution for each pair of blocks of the partition.

We also make use of the concept of left and right cameras; 
we extend the partitioning technique to cameras spaced arbitrarily on the line 
by making use of an intelligent discretization process. The main idea is to partition
the interval containing the left (and right) cameras into a small
number of intervals called ``buckets'', and guess the number of camera pairs between every pair of buckets in the optimal solution. 

In the MAXSUMOFANGLES problem, if we guess that there are $T$ camera pairs involving buckets 
$B$ and $B'$, we cannot simply return $T$ arbitrary pairs involving 
cameras in $B$ and $B'$. Some of these pairs can be too ``sensitive''
for this crude process. Luckily, we get around this difficulty via
a geometric observation about angles which implies that a sensitive pair can be sensitive
at its $B$-end or its $B'$-end but not both. 

For the MINSUMOFRATIOS, the issue of sensitive pairs does not arise, but the main difficulty is that
there is no single ``scale'' of distances at which we can apply the
discretization. (This difficulty would not arise if the ratio
of the maximum inter-camera distance to the minimum inter-camera distance
is polynomially bounded.) So we first apply the discretization at the scale
of the median inter-camera distance (in the optimal solution), and
recursively obtain and solve two independent instances of the problem
with size at most $n/2$. A characterization due to Isler et al.\cite{ikst05}
of the optimal solution for a {\em fixed} camera pairing with the targets
turns out to be quite useful in making this approach work. 

\paragraph{Organization of the paper.} In Section \ref{sec:prelims}, we 
introduce some notation, review some important observations made in
previous work on these problems, and give a high level overview of our algorithms. We present our algorithms for MAXSUMOFANGLES 
and MINSUMOFRATIOS in Sections \ref{sec:maxsum} and \ref{sec:minsum}, 
respectively.

\section{Preliminaries and Notation}
\label{sec:prelims}
Suppose we have some horizontal line $l$ and a set $C$ of $2n$ cameras such that each camera lies on $l$.  We call the $i$th camera on the line $c_i$.  For $c_i, c_j \in C$, we say $c_i < c_j$ if $c_i$ is to the left of $c_j$ on $l$.  We assume that no two cameras have the same position, and thus we have $c_1 < c_2 < \ldots < c_{2n}$.  We are also given a set $T = \{t_1, t_2, \ldots, t_n\}$ of targets that lie in the plane.  Without loss of generality, we can assume that all of the targets lie above $l$.  The distance between cameras $c_i$ and $c_j$ is called the \textit{baseline} of $c_i$ and $c_j$ and is denoted $b_{i,j}$.  A \textit{camera pair} is a set of two distinct cameras, and a \textit{camera pairing} of a set $C'$ of $2m$ cameras is a set of $m$ camera pairs such that each camera in $C'$ appears in exactly one of the camera pairs.

\begin{figure}
\begin{center}
\input{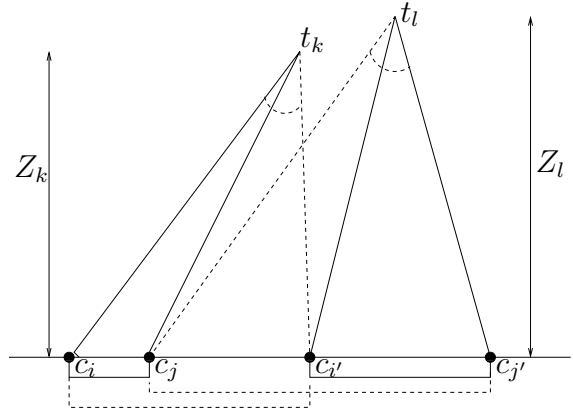}
\caption{For both objective functions, we can do better if we overlap the camera pairs (see dotted lines).}
\label{overlap}
\end{center}

\end{figure}

\begin{figure}
\begin{center}
\input{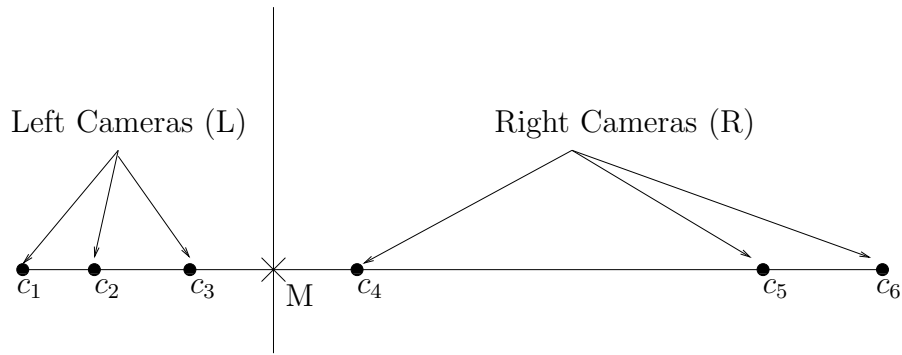}
\caption{In this example $n=3$, cameras $c_1,c_2,c_3$ are ``left'' cameras and the rest are ``right'' cameras. $M$ is the mid-point of baseline $b_{3,4}$ formed by cameras $c_3$ and $c_4$.}
\label{leftrightcams}
\end{center}

\end{figure}

A pairing of cameras is \textit{all-overlapping} if the baselines of any two camera pairs in the pairing intersect.  Suppose there is a camera pairing that contains pairs $(c_i, c_j)$ and $(c_{i'}, c_{j'})$ such that $c_i < c_j < c_{i'} < c_{j'}$ (i.e. the pairing is not all-overlapping).  It is easily seen \cite{gmsvw08} that we can do better for both objective functions if we use the pairs $(c_i, c_{i'})$ and $(c_j, c_{j'})$.  Refer Figure \ref{overlap}. This leads to the observation \cite{gmsvw08} that there is an optimal camera pairing where every camera pair $(c_i, c_j)$ has $i \in \{1, 2, \ldots, n\}$ and $j \in \{n+1, n+2, \ldots, 2n \}$.  Let $M$ denote the midpoint of $b_{n,n+1}$.  See Figure \ref{leftrightcams}. Another way to view this observation is that there is an optimal solution with camera pairing $P$ such that for all pairs $(c_i, c_j) \in P$, $c_i$ is to the left of $M$ and $c_j$ is to the right of $M$.   Our algorithms will only consider such camera pairings.

We shall now present an overview of our algorithms. Although historically the ``Aspect Ratio'' objective function was considered first \cite{ikst05}, it was shown in \cite{gmsvw08} that the ``Tracking Angle'' is in general the most influential way of tracking quality directly. Hence we shall consider this metric first.  Both algorithms use a discretization procedure to aid in computing our solution.  However, a crude discretization procedure will not suffice for either problem, and we must use a more clever technique.  In the case of MAXSUMOFANGLES, a geometric observation comes to our rescue.  In the case of MINSUMOFRATIOS, a more sophisticated discretization process comes to our rescue.

First consider the MAXSUMOFANGLES problem.  Our algorithm constructs a small number of camera ``pairings'' - (a pairing is simply the set formed when each camera is paired with exactly one other camera).  Given this set of camera pairings, we then are able to determine which camera pairing is the best by using a polynomial time algorithm for minimum-weight perfect matching on a bipartite graph in which one set of vertices is the camera pairs \cite{law} and the other is the targets , as observed by Gfeller et al. \cite{gmsvw08}.  

Suppose we fix some optimal solution that uses some all-overlapping camera 
pairing $P$.  That is, every camera pair in $P$ involves one camera to the
left of $M$ and one to the right of $M$.  We discretize the line into a small number of buckets such that the length of each bucket is small compared to its distance from $M$.  Consider a bucket on one side of $M$ and another bucket on the other side of $M$.  Suppose we are able to correctly guess that the number of camera pairs in $P$ that have one camera in each of these buckets is $\mu$.  We would like to be able to arbitrarily pick $\mu$ cameras from the first bucket and pair them with an arbitrary $\mu$ cameras in the second bucket and argue that the tracking angle formed by these camera pairings at the target will not be too small as compared to the one formed by the corresponding pair in $P$.  Unfortunately, this approach will not work, as the arbitrary assignment might cause some tracking angle to be more than an $\epsilon$ factor smaller than the corresponding angle in $P$.  To get around this obstacle, we state and prove a geometric lemma (Lemma \ref{keylemma}) which allows us to handle these ``sensitive'' cases.  The algorithm is then to guess the number of such ``sensitive'' cases that arise for each of our buckets.  We handle those camera pairs (relying heavily on our lemma), and then we use the more crude technique on all of the remaining cameras falling in buckets. If any cameras still remain to be paired, we show that there is an easy way to pair them off (again making use of the same lemma).

Now consider the MINSUMOFRATIOS problem.  Suppose we fix some optimal solution that uses some all-overlapping camera 
pairing $P$.  Let $b^i$ denote the $i$th baseline in $P$ when the baselines are indexed such that $b^1 \leq b^2 \leq \cdots \leq b^n$.  The high level idea of our algorithm is that we will guess the length of the median baseline (i.e. $b^{n/2}$) and then guess all camera pairs in $P$ whose baselines are within a polynomial factor of $b^{n/2}$.  To aid us in guessing, we use a discretization procedure.  This procedure allows us to make all possible guesses in quasi-polynomial time while not doing too much worse than what the optimal solution would have done.  Suppose that we make a correct guess.  Then we will have correctly guessed all camera pairs in $P$ whose baselines fall within some interval $[b^i, b^j]$ for $b^i \leq b^{n/2} \leq b^j$ up to an $\epsilon$ factor.  We then recursively repeat this guessing procedure for baselines in the interval $[b^1, b^{i-1}]$ and the interval $[b^{j+1}, b^{n}]$.  The correctness of the algorithm then follows by showing that the errors do not accumulate too much over the course of the recursion.

Fix a camera pairing $P$ and a set of targets $T$ such that $|P| = |T|$. The
optimal association of camera pairs in $P$ with targets in $T$ is
rather easy to compute, using a characterization given by Isler et al. 
\cite{ikst05}:

\begin{lemma}
 \label{lemma4}
  Let $Z_i$ be the distances of targets in $T$ from line $l$, 
$Z_1 \leq Z_2 \leq \cdots \leq Z_m$ and $b^i$ be the baselines in $P$ sorted such that $b^1 \leq b^2 \leq \cdots \leq b^m$.  There exists an optimal matching 
such that the target at depth $Z_i$ is assigned to the pair with baseline $b^i$.\end{lemma}
We denote the cost of such an assignment of a pairing $P$ to targets $T$ as $\mbox{cost}(P,T)$ and we compute it according to the assignment in Lemma \ref{lemma4}.

\section{Maximizing the Sum of Angles}
\label{sec:maxsum}
We now describe our approximation algorithm for the MAXSUMOFANGLES problem.
We adopt the notation of Section \ref{sec:prelims} -- the camera locations
are $c_1 < c_2 < \cdots < c_{2n}$ on line $l$ , with $c_n < M < c_{n+1}$. The 
targets are points in the plane above the line, and denoted $t_1, \ldots,
t_n$. Let $\epsilon > 0$ be the given approximation parameter; we may assume
that $\eps < 1/2$. Our algorithm outputs a small number of camera pairings of 
$C$. One of the camera pairings that we output will have the property that
it is possible to associate each of the camera pairs in it with targets
in such a way that the sum of the angles at the targets is at least
$(1 - \eps)$ times that of OPT, where OPT is the value of the sum of tracking angles in an optimal solution. For
a given camera pairing, we can compute the best association with the
targets using a polynomial time algorithm for minimum-weight perfect 
matching \cite{law}, as observed by Gfeller et al. \cite{gmsvw08}. We can 
therefore evaluate all of the camera pairings that we output, and return the 
best one.

\begin{figure}
\begin{center}
\input{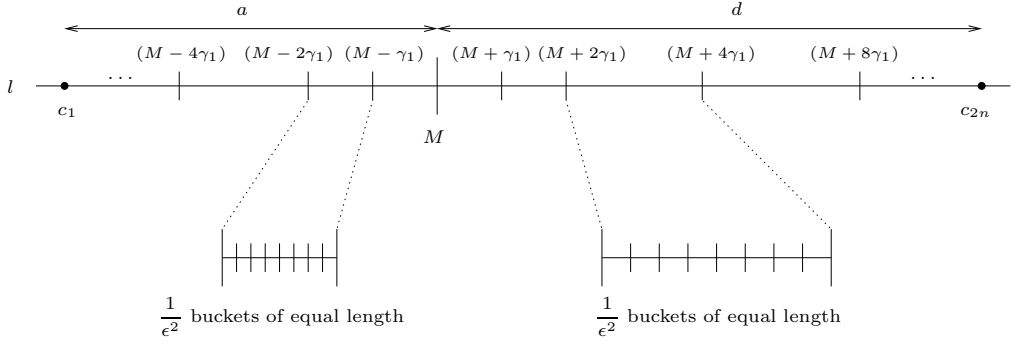}
\caption{Illustration for a conforming partition (the intervals on $l$ are not drawn to scale).  Each of the shown subintervals of $l$ (except for $[M - \gamma_1, M]$ and $[M, M + \gamma_1]$) are divided into $\frac{1}{\epsilon^2}$ buckets of equal length.}
\label{conformingPartition}
\end{center}

\end{figure}

Our algorithm makes use of a subroutine for partitioning an interval to
achieve a result of the following form: Given $0 < \gamma_1 < \gamma_2$,
partition the interval $[M + \gamma_1 , M + \gamma_2]$ 
(resp. $[M - \gamma_2, M - \gamma_1]$ ) into buckets (sub-intervals), so that the distance between the left (resp. right)
endpoint of each bucket from $M$ is at least $\frac{1}{\eps^2}$
times the bucket length. We will call a partition of  
$[M + \gamma_1 , M + \gamma_2]$ (resp. $[M - \gamma_2, M - \gamma_1]$ )
with this property a {\em conforming partition}. Clearly, we
can compute a conforming partition with 
$O\left( \frac{\log (\gamma_2/\gamma_1)}{\eps^2} \right)$ buckets: partition
$[M + \gamma_1, M + 2\gamma_1]$ into $1/\eps^2$ equal sized buckets,
then partition $[M + 2\gamma_1, M + 4\gamma_1]$ into $1/\eps^2$ equal
sized buckets, and so on until we are past $M + \gamma_2$.  We also adopt the
following notation for the algorithm: $L$ will denote the cameras to the
left of $M$, and $R$ the cameras to the right. During the course
of the algorithm, some of the cameras get paired, and $L$ (resp. $R$)
will always stand for the unpaired cameras. Given a
bucket $B$ to the left (resp. right) of $M$, we will let $|B|$ denote the 
number of cameras in the current $L$ (resp. $R$) that fall in bucket $B$.
Finally, let $a = M - c_1$, and $d = c_{2n} - M$. In describing the
algorithm and analysis, we will assume $a \leq d$; the other case
is symmetric.  See Figure \ref{conformingPartition} for an illustration.

\paragraph{The Algorithm.} The following algorithm outputs one candidate
pairing of the cameras $C$ for each combination of the functions
$\sigma$, $\pi$, $\mu$, and $\lambda$ considered. We then evaluate
each of these pairings based on the stated criterion (MAXSUMOFANGLES) and return the best one.

\begin{algorithm*}[ht]
	\caption{}
	\begin{algorithmic}[1]
	\label{algo1}
	
	\STATE Let $B_0$ be the bucket $[M- \frac{\eps a}{100 n^2}, M]$. 
               Let $B_1, \ldots, B_k$ be the buckets resulting from a
               conforming partition of $[M - a, M - \frac{\eps a}{100 n^2}]$.
        \FOR{each map $\sigma: \{B_1, \ldots, B_k\} \rightarrow \Z^+$ 
             such that $\sigma(B_i) \leq |B_i|$}
	\STATE For each $1 \leq i \leq k$, pair the $\sigma(B_i)$ 
               leftmost cameras in $B_i$ with the $\sigma(B_i)$
               leftmost cameras in $R$ arbitrarily.
        \STATE Let $B'_0$ be the bucket $[M, M+ \frac{\eps a}{100 n^2}]$. 
               Let $B'_1, \ldots, B'_j$ be the buckets resulting from a
               conforming partition of $[M + \frac{\eps a}{100 n^2}, M +
               \frac{a}{\eps^2}]$.
        \FOR{each map $\pi: \{B'_1, \ldots, B'_j\} \rightarrow \Z^+$ 
             such that $\sigma(B'_i) \leq |B'_i|$}
         \STATE For each $1 \leq i \leq j$, pair the $\pi(B'_i)$ 
               rightmost cameras in $B'_i$ with the $\pi(B'_i)$
               rightmost cameras in $L$ arbitrarily.
         \FOR{each map $\mu: \{B_0,\ldots,B_k\} \times \{B'_0,\ldots,
                       B'_j\} \rightarrow \Z^+$ such that $\sum_{\ell} 
                       \mu(B_i,B'_{\ell}) \leq |B_i|$ for every $i \leq k$
                       and $\sum_{\ell} \mu(B_{\ell},B'_i) \leq |B'_i|$ for
                       every $i \leq j$}
         \STATE Go through the $(B_i, B'_{\ell})$ pairs in any order
                and arbitrarily pair $\mu(B_i, B'_{\ell})$ cameras
                from $L \cap B_i$ with $\mu(B_i, B'_{\ell})$ cameras
                from $R \cap B'_{\ell}$.
         \FOR{each $0 \leq \lambda \leq |B|$ where $B$ is the bucket 
              $[M - a, M]$}
         \STATE Pair the $\lambda$ leftmost cameras in $L \cap B = L$
                with the $\lambda$ leftmost cameras in $R$ arbitrarily.
         \STATE Pair the remaining cameras in $L$ arbitrarily with cameras
                in $R$.
         \ENDFOR
         \ENDFOR
         \ENDFOR
         \ENDFOR       
\end{algorithmic}
\end{algorithm*}

\paragraph{Running Time.}We will now bound the running time of the algorithm by bounding the number of possibilities that the mappings $\sigma, \pi,$ and $\mu$ consider.  We can bound the number of cameras in any bucket by $n$.  Thus we can bound the number of possibilities for $\sigma$ by $n^{O(k)}$, and we can bound the number of possibilities for $\pi$ by $n^{O(j)}$.  We can bound the number of possibilities for $\mu$ by $n^{O(kj)}$.

Because the number of buckets $j$ and $k$
are bounded by $O\left( \frac{\log (n/\eps)}{\eps^2} \right)$, we have that the running time of the algorithm is bounded by
\[n^{O\left( \frac{\log^2 (n/\eps)}{\eps^4} \right)}.\] This expression
also bounds the number of candidate pairings of $C$ that are output 

\begin{figure}
\begin{center}
\input{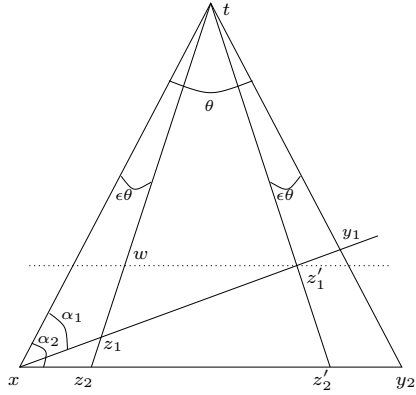}
\caption{Illustration for Lemma \ref{keylemma}.  For $\alpha_2 > \alpha_1$, $\frac{|xz_2'|}{|xz_2|}>\frac{|xz_1'|}{|xz_1|}$.}
\label{fig:f11}
\end{center}
\end{figure}

\paragraph{Approximation Factor.}
Let $\alpha_t$ denote the target angle corresponding to target $t \in T$
in the optimal solution OPT. We now show that there is one choice of
the functions $\sigma$, $\pi$, $\mu$, and $\lambda$ for which it
is possible to associate the candidate pairing output by our algorithm
to the targets in such a way that the target angle $\beta_t$ in this
solution for any target $t$ satisfies
\begin{equation}
\label{eq:key}
 \beta_t \geq  
 \alpha_t - 4 \max \{ \eps \alpha_t, \frac{\eps}{n^2}\sum_t \alpha_t \}.
\end{equation}
Adding this inequality for each $t \in T$ gives $\sum_{t} \beta_t
\geq (1 - 4 \eps) \sum_t \alpha_t$, establishing the algorithm's
correctness. In order to establish this 
inequality, we will find it convenient to move the cameras as part
of the analysis; this will be safe because we only move towards
$M$. For each target $t$, let $l(t)$ and $r(t)$ initially 
denote the cameras associated with $t$ in OPT to the left and right of $M$, 
respectively. As we modify the association of targets with camera
pairs in the analysis, $l(t)$ and $r(t)$ will change. At all times,
$\beta_t$ will denote the angle $l(t) t r(t)$. The following geometric lemma, whose proof we present in the appendix, plays a crucial role in the analysis. Refer Figure \ref{fig:f11}.

\begin{lemma}
 \label{keylemma}
Let $x,y,t$ be three non-collinear points in the plane such that $\myangle{xty}=\theta$. Let $z$ and $z'$ be two points on the line $xy$ such that both $z$ and $z'$ lie between $x$ and $y$ and 
$\myangle{xtz}=\myangle{ytz'}=\epsilon \cdot \theta$. Recall that
$0 < \eps < 1/2$. Then the ratio $\frac{|xz'|}{|xz|}\le \frac{1}{\epsilon^2}$ 
where $|xz|,|xz'|$ are the lengths of segments $xz, xz'$ respectively.
\end{lemma}

\paragraph{Fixing $\sigma$.} In this step, let us call a target
$t$ {\em sensitive} if $l(t)$ lies in bucket $B_i$ for $i \geq 1$
(so not in the first bucket $B_0$), and the angle
$l(t) t p$ is at least $\eps$ times the angle $\beta_t = l(t) t r(t)$,
where $p$  is the right endpoint of bucket $B_i$. Since the
buckets come from a conforming partition, Lemma \ref{keylemma}
implies that for a sensitive $t$ the angle $l(t) t c$ for any
camera $c$ to the right of $M$ is at least $(1 - \eps) \beta_t$. (The
upshot is that we have to be careful in changing $l(t)$, but
we have considerable flexibility with $r(t)$.) Fix $\sigma$
so that $\sigma(B_i)$ is the number of sensitive targets $t$ with
left endpoints within $B_i$. With this choice of $\sigma$, recall
that our algorithm fixes $\sum_i \sigma(B_i)$ camera pairs at this
stage. Now for each sensitive $t$, reset $l(t)$ so that it lies
to the left of (or is the same as) the original $l(t)$ and is one of the 
$\sigma(B_i)$ leftmost points in its bucket $B_i$, and reset $r(t)$ to be
the camera that is paired with the new $l(t)$ in the partial camera
pairing that is fixed. It follows that now $\beta_t \geq (1 - \eps)
\alpha_t \geq \alpha_t - \max \{\eps \alpha_t,\frac{\eps}{n^2} \sum_t \alpha_t \}$ for all sensitive $t$. The association of sensitive
$t$ with the camera pairing we output is finalized now, and hence
Inequality \ref{eq:key} holds for such $t$. 

For the $t$ that are not sensitive, reset $l(t)$ to be some other
camera in the same bucket as the original $l(t)$ and reset $r(t)$
to be a camera that is the same or to the right of the original $r(t)$.
(During these resettings, we always ensure that each camera is $l(t)$
or $r(t)$ for exactly one $t$.) Now ``move'' the camera that is $l(t)$
for each such $t$ to the right endpoint of the bucket containing
$l(t)$.

\begin{figure}
\begin{center}
\input{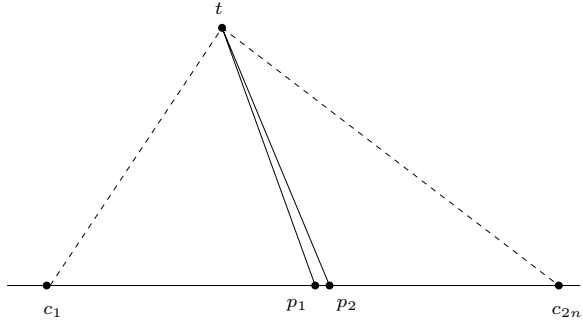}
\caption{A figure to illustrate that angle($p_1tp_2$) is very small compared to angle($c_1tc_{2n}$).}
\label{smallAngle}
\end{center}

\end{figure}

For a $t$ that is not sensitive and for which $l(t)$ does not lie
in bucket $B_0$, it is clear that the new $\beta_t$ is at least
$(1 - \eps)$ times the original $\beta_t$. Now if $l(t)$ lies
in $B_0$, letting $p_1$ and $p_2$ denote the left and right
endpoints of $B_0$, we argue that now
\[ \beta_t \geq \alpha_t - \myangle{p_1 t p_2} \geq
    \alpha_t - \frac{\eps}{n^2} \myangle{c_1 t c_{2n}} \geq
    \alpha_t - \frac{\eps}{n^2} \sum_t \alpha_t. \]
For the second inequality, we show that $ \myangle{p_1 t p_2} 
\leq \frac{\eps}{n^2} \myangle{c_1 t c_{2n}}$; we omit from
this version the somewhat tedious argument for this claim,
but note that it is here that we use the fact that $t$ is
outside the circle with diameter $c_1 c_{2n}$.  See Figure \ref{smallAngle} for a figure to provide intuition for why the claim holds.

Thus, for the $t$ that are not sensitive, we have
\[ \beta_t \geq \alpha_t - \max \{ \eps \alpha_t, 
\frac{\eps}{n^2} \sum_t \alpha_t \}.\]

We move on to the next stage with the targets $t$ whose association with
camera pairings has not been finalized 
(the targets that were not sensitive in this stage), along with the
cameras $l(t)$ and $r(t)$ for such targets (the cameras that have
not been paired up by our algorithm).

\paragraph{Fixing $\pi$.} This is quite symmetric to the previous step.
Among the targets that have
moved on to this step, let us call a target
$t$ {\em sensitive} if $r(t)$ lies in bucket $B'_i$ for $i \geq 1$
(so not in the first bucket $B'_0$), and the angle
$p t r(t)$ is at least $\eps$ times the angle $\beta_t = l(t) t r(t)$,
where $p$  is the left endpoint of bucket $B'_i$. Since the
buckets come from a conforming partition, Lemma \ref{keylemma}
implies that for a sensitive $t$ the angle $c t r(t)$ for any
camera $c$ to the left of $M$ is at least $(1 - \eps) \beta_t$. Fix $\pi$
so that $\pi(B'_i)$ is the number of sensitive targets $t$ with
$r(t)$ within $B'_i$. With this choice of $\pi$, recall
that our algorithm now fixes $\sum_i \pi(B'_i)$ camera pairs at this
stage. Now for each sensitive $t$, reset $r(t)$ so that it lies
to the right of (or is the same as) the original $r(t)$ and is one of the 
$\pi(B'_i)$ rightmost points in its bucket $B'_i$, and reset $l(t)$ to be
the camera that is paired with the new $r(t)$ in the partial camera
pairing that is now fixed. It follows that for all sensitive $t$,
the current $\beta_t$ is at least $(1 -\eps)$ times the $\beta_t$
at the end of the previous step, and thus
$\beta_t \geq \alpha_t - 2 \max \{\eps \alpha_t,\frac{\eps}{n^2} \sum_t \alpha_t \}$. The association of sensitive
$t$ with the camera pairing we output is finalized now, and hence
Inequality \ref{eq:key} holds for such $t$. 

For the $t$ that are not sensitive, reset $r(t)$ to be some other
camera in the same bucket as the original $r(t)$ and reset $l(t)$
to be a camera that is the same or to the left of the original $l(t)$.
Now ``move'' the camera that is $r(t)$
for each such $t$ to the left endpoint of the bucket containing $r(t)$.

For a $t$ that is not sensitive, we can argue along the
same lines as in the previous step to conclude that
$\beta_t \geq \alpha_t - 2 \max \{ \eps \alpha_t, \frac{\eps}{n^2} 
\sum_t \alpha_t \}$.

We move on to the next stage with the targets $t$ whose association with
camera pairings has not been finalized 
(the targets that were not sensitive in this stage), along with the
cameras $l(t)$ and $r(t)$ for such targets (the cameras that have
not been paired up by our algorithm).

\paragraph{Fixing $\mu$.} For
every pair $0 \leq i \leq k$ and $0 \leq i' \leq j$, fix
$\mu(i,i')$ to be the number of surviving targets $t$ with
$l(t) \in B_i$ and $r(t) \in B'_{i'}$. With this choice of
$\mu$, recall that our algorithm outputs $\mu(i,i')$ pairs
of cameras at this stage with one endpoint in $B_i$
and the other in $B'_{i'}$. Reassign the $l(t)$
and $r(t)$ values for the surviving targets 
(this reassignment should not cause an $l(t)$ or $r(t)$ value
to move to a different bucket), if necessary, so that these
$\mu(i,i')$ pairs that are output by the algorithm are
assigned to the targets that led to the definition of
$\mu(i,i')$. The association of such targets with the
camera pairing that we output is finalized at this stage.
Notice that $\beta_t$ does not change for any target in
this step, whether it is finalized or not. This is because
the cameras have already been moved to endpoints of the
bucket they belong to. In particular, this means that 
Inequality \ref{eq:key} holds for the targets that do not
survive this stage.

We now move to the final stage with the targets that 
survive. Note that the $l(t)$ for such targets 
belong to the bucket $B = [M-a, M]$, and the $r(t)
\geq M + \frac{a}{\eps^2}$.

\paragraph{Fixing $\lambda$.} In this step, call a surviving
target $t$ sensitive if $\myangle{l(t) t M} \geq
\eps \cdot \myangle{l(t) t r(t)} = \eps \beta_t$. Lemma \ref{keylemma}
implies that for a sensitive $t$ the angle $l(t) t c$ for any
camera $c$ to the right of $M + \frac{a}{\eps^2}$ is at least 
$(1 - \eps) \beta_t$. Fix $\lambda$ to be the number of such
sensitive targets $t$. With this choice of $\lambda$, recall
that our algorithm fixes $\lambda$ camera pairs at this
stage. Now for each sensitive $t$, reset $l(t)$ so that it lies
to the left of (or is the same as) the original $l(t)$ and is one of the 
$\lambda$ leftmost points in its bucket $B$, and reset $r(t)$ to be
the camera that is paired with the new $l(t)$ in the partial camera
pairing that is fixed. It follows that for sensitive $t$,
the new $\beta_t$ is at least $(1 - \eps)$ times the $\beta_t$ before
this step, and thus
$\beta_t \geq \alpha_t - 3 \max \{\eps \alpha_t,\frac{\eps}{n^2} \sum_t \alpha_t \}$. The association of sensitive
$t$ with the camera pairing we output is finalized now, and hence
Inequality \ref{eq:key} holds for such $t$. 

For the $t$ that are not sensitive, reset $l(t)$ to be some other
camera in $B$, and reset $r(t)$ to be a camera that is the same or to the 
right of the original $r(t)$. Move the camera $l(t)$ to the point $M$.
The new $\beta_t$ is at least $(1 - \eps)$ times the $\beta_t$ before
this step, and thus 
$\beta_t \geq \alpha_t - 3 \max \{\eps \alpha_t,\frac{\eps}{n^2} \sum_t \alpha_t \}$.

By reassigning $l(t)$, if necessary, but not the $r(t)$, the camera pairs that 
the algorithm outputs now will have the form $(l(t), r(t))$ for each 
surviving $t$.
This completes the association for all the targets.
This reassignment does not change $\beta_t$ because each $l(t)$ is on
point $M$, and Inequality \ref{eq:key} holds for every $t$.

This completes our argument about the approximation guarantee.

\begin{theorem}
 There is an algorithm for the MAXSUMOFANGLES problem that, for any
$0 < \eps < 1$, runs in quasi-polynomial time and returns a solution in which the sum of tracking angles is at least $(1 - \eps)$ times that in the optimal solution.
\end{theorem}

\section{Minimizing the Sum of Ratios}
\label{sec:minsum}

We now present our algorithm for the MINSUMOFRATIOS problem. Recall that
in this problem the cost of assigning a camera pair $(c_i, c_j)$ to target $t_k$ is $\frac{Z_k}{b_{i,j}}$ where $Z_k$ is the vertical distance between $t_k$ and the line  $l$ containing the cameras, and $b_{i,j}$ is the baseline corresponding to (distance between) $c_i$ and $c_j$.  We would like to minimize the sum of these costs.  In this section, we present an algorithm that returns a solution whose cost (sum of aspect ratios) is no worse than $(1 + \epsilon)$ times the same in the optimal solution, in quasi-polynomial time for any $\epsilon > 0$.  

\paragraph{The Algorithm.}

Our algorithm is recursive and is formally defined as Algorithm \ref{algo2}. 
The algorithm takes as input a set of $2m$ cameras $X$, a set of $m$ targets $Y$, and two positive real numbers $\ell$ and $u$. The cameras are indexed 
$x_1, x_2, \ldots x_{2m}$ such that $x_1 < x_2 < \cdots < x_{2m}$ and
$x_m < M < x_{m+1}$ and the targets are indexed $y_1, y_2, \ldots y_m$ such that $Z_1 \leq Z_2 \leq \cdots \leq Z_m$.  The algorithm will return a camera pairing $P$ of the cameras in $X$ such that the baselines of all of the pairs in $P$ are at least $\ell$ and are at most $u$.  If no such pairing is possible,
our algorithm returns a ``dummy'' camera pairing $I$;  for ease of
description, we define  $\mbox{cost}(P,T) = \infty$ for any pairing
$P$ that contains $I$. Our algorithm breaks up the line $l$ into intervals
called buckets. We will use $|d|$ to denote the number of cameras in
$X$ that lie in bucket $d$. We assume $0 < \eps < 1/2$.


\begin{algorithm*}[ht]
	\caption{$\mbox{minRatioPair}(X,Y,\ell,u)$}
	\begin{algorithmic}[1]
	\label{algo2}
	
	\STATE If $|X| = 0$, return the empty set.
	\STATE Let $\mathcal{L}$ be the set of all baselines $b_{i,j}$ such that 
         $1 \leq i \leq m$, $m+1 \leq j \leq 2m$, and 
         $\ell \leq b_{i,j} \leq u$.
	\STATE Set local variable $best \leftarrow \infty$, and 
         $P_{best} \leftarrow I$.

	\FOR{each $\beta \in \mathcal{L}$}
		\STATE Partition the interval $[M - 2n\beta, M]$ into
                $O(\frac{\log n}{\epsilon})$  buckets in the following way.
                Initialize $r = M$.  

		\FOR{$i=-1, 0, 1, \ldots, 2\log n$}
			\STATE Discretize the interval $[r-2^i\frac{\beta}{n},r]$ into $\frac{2}{\epsilon}$ buckets of equal length.  
			\STATE $r \leftarrow 2^i\frac{\beta}{n}$
		\ENDFOR
		\STATE Symmetrically partition the interval $[M, M + 2n \beta]$
                       into $O(\frac{\log n}{\epsilon})$  buckets.
		\STATE Let $\mathcal{B}_1$ denote the set of all buckets to the left of 
                $M$, and let $\mathcal{B}_2$ denote the set of buckets to the right of $M$.               \FOR{each map $\mu: \mathcal{B}_1 \times \mathcal{B}_2 \rightarrow \Z^+$ and each
                     map $\sigma: \mathcal{B}_1 \cup \mathcal{B}_2 \rightarrow \Z^+$ such that
                     $\sigma(B) + \sum_{B' \in \mathcal{B}_2} \mu(B,B') \leq |B|$ for each
                     $B \in \mathcal{B}_1$, $\sigma(B') + \sum_{B \in \mathcal{B}_1} \mu(B,B') 
                     \leq |B'|$ for each $B' \in \mathcal{B}_2$, and $\sum_{B \in \mathcal{B}_1}
                     \sigma(B) = \sum_{B' \in \mathcal{B}_2} \sigma(B')$}
				
		\STATE $X_{short} \leftarrow \emptyset, X_{mid} \leftarrow \emptyset, X_{long} \leftarrow \emptyset$, $P_{mid} \leftarrow \emptyset$

				\STATE Go through each $(B,B') \in \mathcal{B}_1 \times
\mathcal{B}_2$ in any order and pair the $\mu(B,B')$ cameras in $B$ that are closest 
to $M$ with the $\mu(B,B')$ cameras in $B'$ that are closest to $M$ and 
place these pairs in $P_{mid}$.  Add these cameras to $X_{mid}$ and remove them from 
their respective buckets.

\STATE For each bucket $B \in \mathcal{B}_1 \cup \mathcal{B}_2$: Of the cameras that remain in 
$B$, place the $\sigma(B)$ cameras that are farthest from $M$ into $X_{short}$.  Place all other remaining cameras in $B$ into $X_{long}$.

\STATE Denoting $m_s \equiv \frac{|X_{short}|}{2}$ and $m_{mid} \equiv
       \frac{|X_{mid}|}{2}$,  let 
       $Y_{short} = \{y_1, y_2, \ldots, y_{m_s}\}$, 
$Y_{mid} = \{ y_{m_s +1}, y_{m_s + 2}, 
\ldots, y_{m_s + m_{mid}}\}$, and $Y_{long} = \{y_{m_s + m_{mid} + 1}, 
\ldots, y_m\}$.

\IF{ $|Y_{short}| \leq m/2$ and $|Y_{long}| \leq m/2$ }
\STATE $P \leftarrow P_{mid} \cup \mbox{minRatioPair}(X_{short}, Y_{short}, \ell, (1+\epsilon)\frac{\beta}{2n}) \cup \mbox{minRatioPair}(X_{long}, Y_{long}, 
(1 - \epsilon)2n\beta, u (1 + \eps))$

				\STATE If cost$(P,Y) < best$, then $best \leftarrow \mbox{cost}(P,Y)$ and $P_{best} \leftarrow P$.
\ENDIF

\ENDFOR		

\ENDFOR 

	\STATE Return $P_{best}$.
	
	\end{algorithmic}
\end{algorithm*}

\begin{figure}
\begin{center}
\input{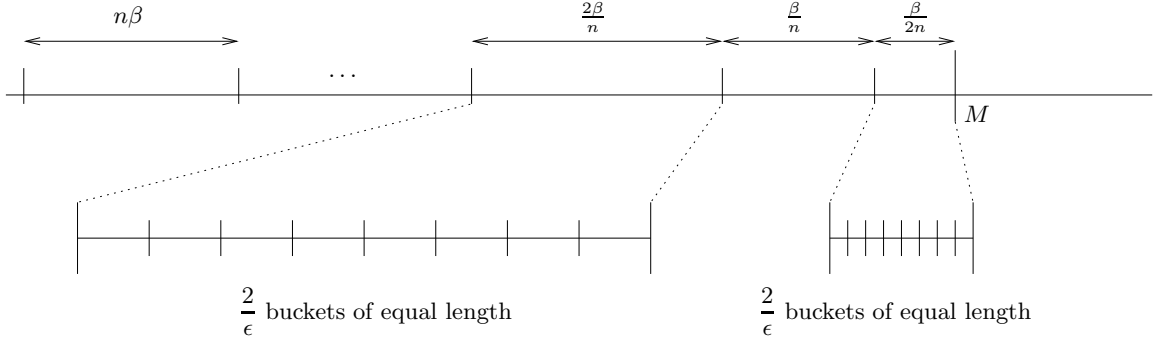}
\caption{Illustration for the discretization process.  The value in each interval 
denotes the length of the interval.  Each interval is divided into 
$\frac{2}{\epsilon}$ buckets of equal length.}
\label{d2}
\end{center}

\end{figure}

\begin{figure}[htpb]
\begin{center}
 \input{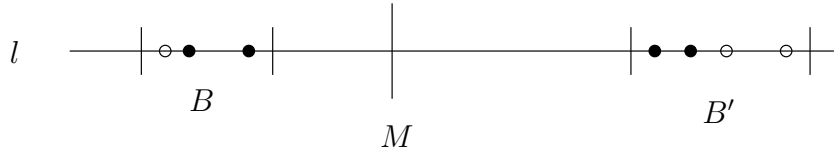}

\caption{Illustration for the step in algorithm where the bucket pair 
$(B,B')$ contributes to $X_{mid}$. The filled in cameras will be paired up and 
placed into $X_{mid}$ for $\mu(B,B') = 2$}
\label{planargadget}
\end{center}

\end{figure}

For solving the input instance of MINSUMOFRATIOS, we invoke
$\mbox{minRatioPair}(C,T,b_{n,n+1},b_{1,2n})$. For this invocation, note
that the depth of the recursion is at most $\log n$, because the
size of $Y$ falls by a factor of $2$ with each recursive call. It will
also be useful to note that when we make a recursive call, the
lower bound $\ell$ for the recursive call is not smaller than the original
$\ell$; the upper bound $u$ for the recursive call can be larger than
the original, but only by a factor of $(1 + \eps)$.

\paragraph{Running Time.}
We will now show that Algorithm \ref{algo2} runs in quasi-polynomial time
when the instance $\mbox{minRatioPair}(C,T,b_{n,n+1},b_{1,2n})$ is
invoked. The set $\mathcal{L}$ contains $O(m^2)$ elements.  The number of buckets in
$\mathcal{B}_1 \cup \mathcal{B}_2$ is $O(\frac{\log n}{\epsilon})$. This means that
the number of choices of $\mu$ and $\sigma$ is
at most $(m + 1)^{O( \frac{\log^2 n}{\epsilon^2})}$. Thus we make
$(m + 1)^{O( \frac{\log^2 n}{\epsilon^2})} = 
n^{O( \frac{\log^2 n}{\epsilon^2})}$ direct recursive calls. Since 
the depth of the recursions is at most $\log n$, this gives
us our quasi-polynomial running time.

\paragraph{Approximation Ratio.}
We now give an argument for the approximation factor guaranteed which, though
informal, highlights the main issues. Consider the optimal pairing
$P_{OPT}$ for the input instance with cameras $C$ and targets $T$, and the
optimal association of targets in $T$ with cameras in $P_{OPT}$. We show
we can associate each target $t$ with a camera pair in such a way that
(a) these camera pairs form a camera pairing of $C$, (b) the baseline
of the camera pair associated with each $t$ is at least $(1 - 5\eps)$
times the corresponding baseline in $P_{OPT}$, and (c) our algorithm
outputs a solution that is at least as good as this association. Thus,
we obtain our $(1 + O(\eps))$ approximation factor.

Let us describe how this special association is constructed. Let
us start with the optimal association of each $t \in T$ with the corresponding 
camera pair in $P_{OPT}$, and consider the invocation
$\mbox{minRatioPair}(C,T,b_{n,n+1},b_{1,2n})$. Consider the situation
where the algorithm chooses $\beta$ to be the median baseline in
$P_{OPT}$. With this $\beta$ it computes bucket sets $\mathcal{B}_1$ and $\mathcal{B}_2$.
The pairs in $P_{OPT}$ can be split into three sets: those whose
baselines are strictly smaller than $\beta/2n$ (the short baselines), 
those whose baselines are between $\beta/2n$ and $2n\beta$ (the medium
baselines), and those whose baselines are strictly larger
than $2n\beta$ (the long baselines). Consider the algorithm's
choice of $\mu$ so that $\mu(B,B')$ equals the number of medium
baselines with endpoints in $B$ and $B'$, and the choice of $\sigma$
so that $\sigma(B)$ equals the number of cameras in $B$ that are
endpoints of short baselines. Notice that with this choice of
$\mu$ the algorithm constructs sets $Y_{short}$, $Y_{mid}$, and
$Y_{long}$ of sizes $m_s = \sum_B \sigma(B)$, 
$m_{mid} = \sum_{(B,B')} \mu(B,B')$, and $m - m_s - m_{mid}$, respectively, 
and sets $X_{short}$, $X_{mid}$, and $X_{long}$ of sizes 
$2 m_s$, $2 m_{mid}$, and $2 (m - m_s - m_{mid})$, respectively.
It also constructs a pairing $P_{mid}$ of the points in $X_{mid}$. 
We show that from $P_{OPT}$ we can obtain also a pairing $P_{short}$
of $X_{short}$ and a pairing $P_{long}$ of $X_{long}$, and 
modify our initial association of targets
so that:

\begin{enumerate}
\item Each target in $Y_{mid}$ is associated with a pair in $P_{mid}$
which is at most $(1 - \eps)$ times shorter than the pair it
was initially associated with.
\item Each target in $Y_{short}$ is associated with a pair
in $P_{short}$ that is at least as long as the pair it was associated
with.
\item Each target in $Y_{long}$ is associated with a pair
in $P_{long}$ that is shorter than the original pair it
was associated with {\em by at most an additive} $\eps 2 n \beta$.
\end{enumerate}

(In addition, baselines in $P_{short}$ have length at most
$(1 + \eps) \frac{\beta}{2n}$, and this explains the
upper bound in the recursive call 
$\mbox{minRatioPair}(X_{short}, Y_{short}, \ell, (1+\epsilon)\frac{\beta}{2n})$.
Similarly for the other recursive call.)

At this point, we have finalized our special association for targets in
$Y_{mid}$. This is the association in (1) above; notice that
we lose only a $(1 - \eps)$ factor.
For the targets in $Y_{short}$, we
``recursively'' construct the special association in this way
starting with the new association with $P_{short}$ and following
the recursive call $\mbox{minRatioPair}(X_{short}, Y_{short}, \ell, (1+\epsilon)\frac{\beta}{2n})$. Similarly for the targets in $Y_{long}$.

Notice that the baselines associated with targets in $Y_{short}$ 
have not shrunk; the baselines associated with targets in
$Y_{long}$ may have shrunk, but by at most an additive
$\eps 2 n \beta$. Now for a given target $t$, how much shrinkage
can it experience as we recursively construct our
special association?  This is no more than the sum of
the $\eps 2 n \beta$ terms over all the $\beta$'s that it
sees and for which it is in $Y_{long}$. Now, Lemma
\ref{factor2} implies that this can be bounded by a
geometric series that is at most $4 \eps n \beta$ for
the largest $\beta$ it is in $Y_{long}$ for. And this
in turn means that in our special association, 
$t$ is associated with a baseline that is at least
$(1 - 5 \eps)$ times the corresponding baseline in $P_{OPT}$.

\begin{lemma}
Consider two recursive calls made by the algorithm where the
second recursive call is nested (possibly by several levels) within
the first. Let $\beta_1$ denote the choice of $\beta$ in the
first call within which the second call is contained, and let
$\beta_2$ denote any choice of $\beta$ within the second call.
Then either $\beta_2 \geq 2 \beta_1$ or $\beta_2 \leq \beta_1/2$.
\label{factor2}
\end{lemma}

\begin{proof}
The immediate recursive calls that we make from the first call with
$\beta_1$ either have an upper bound of $(1 + \eps) \frac{\beta_1}{2n}$
or a lower bound of $(1 - \eps) 2 n \beta_1$. Now lower bounds do
not decrease with recursion in our algorithm, so $\beta_2 > ( 1 - \eps)
2n \beta_1 \geq 2 \beta_1$ ($\eps < 1/2$) if the second call is nested
within an immediate call of the latter type. Upper bounds may
increase with recursion but only by a factor of $(1 + \eps)$, and
since the depth of recursion is less than $\log n$, we have
$\beta_2 \leq (1 + \eps)^{\log n} \frac{\beta_1}{2n} \leq \beta_1/2$
if the second call is nested within an immediate call of the former type.
\end{proof}

\begin{theorem}
There is an algorithm for MINSUMOFRATIOS that, for any parameter
$0 < \eps < 1$, runs in quasi-polynomial time and 
returns a solution whose cost is at most $(1 + \eps)$ times that
of the optimal solution.
\end{theorem}

\bibliography{cameras}

\begin{thebibliography}{1}

\bibitem{ah97}
Esther~M. Arkin and Refael Hassin.
\newblock On local search for weighted k-set packing.
\newblock In Rainer~E. Burkard and Gerhard~J. Woeginger, editors, {\em ESA},
  volume 1284 of {\em Lecture Notes in Computer Science}, pages 13--22.
  Springer, 1997.

\bibitem{gmsvw08}
Beat Gfeller, Mat{\'u}s Mihal{\'a}k, Subhash Suri, Elias Vicari, and Peter
  Widmayer.
\newblock Angle optimization in target tracking.
\newblock In Joachim Gudmundsson, editor, {\em SWAT}, volume 5124 of {\em
  Lecture Notes in Computer Science}, pages 65--76. Springer, 2008.

\bibitem{hz00}
R.~I. Hartley and A.~Zisserman.
\newblock {\em Multiple View Geometry in Computer Vision}.
\newblock Cambridge University Press, ISBN: 0521623049, 2000.

\bibitem{ikst05}
Volkan Isler, Sanjeev Khanna, John~R. Spletzer, and Camillo~J. Taylor.
\newblock Target tracking with distributed sensors: The focus of attention
  problem.
\newblock {\em Computer Vision and Image Understanding}, 100(1-2):225--247,
  2005.

\bibitem{law}
Eugene Lawler.
\newblock {\em Combinatorial Optimization: Networks and Matroids}.
\newblock Dover Publications (Originally published in 1976), 2001.

\bibitem{s00}
Frits C.~R. Spieksma.
\newblock Chapter 1: Multi index assignment problems: Complexity,
  approximation, applications.

\bibitem{sw96}
Frits C.~R. Spieksma and Gerhard~J. Woeginger.
\newblock Geometric three-dimensional assignment problems.
\newblock {\em European Journal of Operational Research}, 91(3):611--618, June
  1996.

\bibitem{yjs06}
Alper Yilmaz, Omar Javed, and Mubarak Shah.
\newblock Object tracking: A survey.
\newblock {\em ACM Comput. Surv.}, 38(4):13, 2006.

\end{thebibliography}

\appendix

\section{Proof of Lemma \ref{keylemma}}
 Denote $\myangle{txy}$ by $\alpha$. For greater values of $\alpha$, the rays $\stackrel{\longrightarrow}{ty}$ and $\stackrel{\longrightarrow}{xy}$ meet further apart and in the limiting case i.e. for $\alpha=\pi - \theta$ these become parallel. Thus, we are interested in the range $0<\alpha<\pi - \theta$. First we shall show that within this range as $\alpha$ increases, the ratio $\frac{|xz'|}{|xz|}$ strictly increases. Then we shall show that in the limiting case - i.e. when $\alpha \rightarrow \pi - \theta$, the ratio $\frac{|xz'|}{|xz|}\rightarrow \frac{1}{\epsilon^2}$. This will complete the proof.

For the first part, refer back to Figure \ref{fig:f11}. Here we consider two different values of $\alpha$ such that $0<\alpha_1 <\alpha_2 < \pi - \theta$. Let $y_1,z_1,z_1'$ and $y_2,z_2,z_2'$ denote the corresponding points as defined before for these two cases.

Now consider the line through $z_1'$ that is parallel to $xy_2$. Let this line intersect the line $tz_2$ at point $w$. Now clearly $\triangle{z_1xz_2}$ and $\triangle{z_1z_1'w}$ are similar since two of their angles are equal. Hence, $\frac{|z_1'w|}{|xz_2|}=\frac{|z_1z_1'|}{|z_1x|}$. But clearly $|z_2z_2'|>|z_1'w|$. Hence, we have $\frac{|z_2z_2'|}{|xz_2|}>\frac{|z_1z_1'|}{|xz_1|}$. This clearly implies $\frac{|xz_2'|}{|xz_2|}>\frac{|xz_1'|}{|xz_1|}$. Thus as $\alpha$ increases our ratio increases.

\begin{figure}
\begin{center}
\input{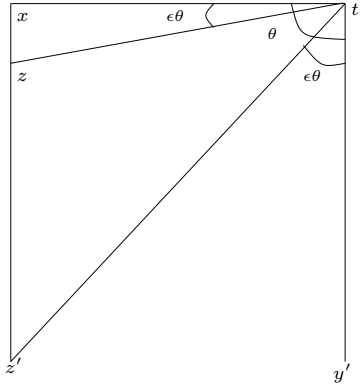}
\caption{Illustration for Lemma \ref{keylemma}.  In the limiting case, the ratio $\frac{|xz'|}{|xz|}$ is at most $\frac{1}{\epsilon^2}$.}
\label{fig:f2}
\end{center}
\end{figure}

For the second part, refer to Figure \ref{fig:f2}. We consider the limiting 
case in 
which $\alpha=\pi-\theta$. As we observed before in this case rays $\stackrel{\longrightarrow}{ty}$ and $\stackrel{\longrightarrow}{xy}$ become parallel.
Let $y'$ be the point on $ty$ at a distance $|xz'|$ from $t$, so that we
have the parallelogram $txz'y'$. In this scenario, $\triangle{txz}$ is similar to $\triangle{ty'z'}$ (The Figure is not to scale). This is because $\myangle{xtz}=\myangle{y'tz'}=\epsilon \cdot \theta$ and $\myangle{txz}=\myangle{ty'z'}$ 
since these are opposite angles of parallelogram $txz'y'$. Now, due to similarity of triangles $\triangle{txz}$ and $\triangle{ty'z'}$, $\frac{|xz|}{|y'z'|}=\frac{|tx|}{|ty'|}$. Hence, $|ty'|=\frac{|tx||y'z'|}{|xz|}$. But note that $|ty'|=|xz'|$ and $|y'z'|=|tx|$ since these are opposite sides of the parallelogram.

After plugging in these values we have $\frac{|xz'|}{|xz|}=\frac{|tx|^2}{|xz||xz|}=\frac{|tx|^2}{|xz|^2}$. By applying the sine rule in $\triangle{txz}$, we have $\frac{|tx|}{\sin{\myangle{tzx}}}=\frac{|xz|}{\sin{\myangle{xtz}}}$. But note that $\myangle{xtz}=\epsilon\cdot\theta$ and $\myangle{tzx}=\myangle{zty}=(1-\epsilon)\cdot\theta$ since these form a pair of alternate angles between parallel lines $ty'$ and $xz'$ cut by transversal $tz$.

Hence we have $(\frac{|tx|}{|xz|})^2=(\frac{\sin{(1-\epsilon)\theta}}{\sin{\epsilon\cdot\theta}})^2\le(\frac{1-\epsilon}{\epsilon})^2\le\frac{1}{\epsilon^2}$. 

Thus $\frac{|xz'|}{|xz|}\le \frac{1}{\epsilon^2}$

\end{document}